# ROSAT HRI observations of IRAS P09104+4109: a massive cooling flow

A.C. Fabian and C.S. Crawford
*Institute of Astronomy, Madingley Road, Cambridge CB3 0HA*

**ABSTRACT**
A ROSAT HRI image of the hyperluminous infrared galaxy IRAS P09104+4109 shows extended X-ray emission peaked around the central galaxy in a profile characteristic of a cooling flow. The inferred soft X-ray luminosity is comparable to that detected with ASCA. The ASCA spectrum is well fit by a cooling flow spectrum, indicating a mass cooling rate of about $1000\,{\rm M}_\odot\,{\rm yr}^{-1}$. The cooling flow dominates the total luminosity of the surrounding cluster. Any contribution from a hidden AGN in the host galaxy to either the total ASCA spectrum or the HRI image is small.

## 1 INTRODUCTION

The infrared-luminous galaxy IRAS P09104+4109 at redshift $z = 0.442$ (Kleinmann et al 1988) has recently been shown to be a powerful X-ray source with a rest frame luminosity in the 2–10 keV band of $2 \times 10^{45}\,{\rm erg\,s^{-1}}$ (Fabian et al 1994a; $H_0 = 50\,{\rm km\,s^{-1}\,Mpc^{-1}}$ is assumed throughout). The optical spectrum has strong, narrow emission lines (Kleinmann et al 1988) and a highly polarized continuum, indicating that much of the light is scattered (Hines & Wills 1993). The general impression, supported by the detection of a broad MgII emission line, is that IRAS P09104+4109 contains a powerful, obscured, active nucleus (Hine & Wills 1993). The cD host galaxy lies in the centre of a flattened cluster of galaxies (Kleinmann et al 1988).

The ASCA X-ray spectrum of IRAS P09104+4109 appeared to support the active nucleus hypothesis since it is well fitted by a power-law of photon index $\Gamma \approx 2$, similar to that of many AGN. In addition, a strong emission line identified as due to helium-like iron was observed. Such a line is expected due to recombination and resonance scattering if the nucleus is hidden and the X-ray continuum scattered into our line of sight by electrons in the cooling intracluster medium within about a kpc from the nucleus (Fabian et al 1994a). The X-ray emission is consistent, at the resolution of ASCA ($FWHM \approx 50$ arcsec) with a point source; it is certainly much less extended than the more distant cluster CL 0016+16.

In this Letter we report a ROSAT High Resolution Imager (HRI) observation of IRAS P09104+4109 which resolves the X-ray source spatially and demonstrates that the bulk of the X-ray emission is from the intracluster gas and is not scattered nucleus emission. The X-rays show the highly-peaked profile of a massive cooling flow around the host cD galaxy.

## 2 THE ROSAT HRI IMAGE

IRAS P09104+4109 was observed with the ROSAT HRI for 7937 s on 1994 November 8. An image of the source binned into 1 arcsec pixels is shown in Fig. 1 and a contour map in Fig. 2. If the source were pointlike, then 50 per cent of the counts from the source would lie within a radius of 3 arcsec (David et al 1993). The source is clearly much more extended with 50 per cent of the counts beyond 14 arcsec. The total extent is difficult to establish since the signal-to-noise ratio drops rapidly at larger radii; it clearly exceeds several tens of arcsec (Fig. 3). We detect $368 \pm 28$ counts within a radius of 100 arcsec after correcting for background.

Curiously the source appears to have a hole at the centre with a bright surrounding ring. The width of the hole, of about 4 arcsec, is consistent with the point response function of the HRI and so should be real. The aspect-error housekeeping data indicate pointing uncertainties of less than a few arcsec and the hole is apparent in both halves of the data when split in time. It is difficult to estimate the significance of this structure since the real profile of the centre of the source is unknown. We have also examined the second brightest source in the field which unfortunately only has about 15 ct. These appear confined to a region of size similar to the point spread function (PSF) of the HRI, indicating that the HRI image is stable.

The X-ray luminosity of IRAS P09104+4109 in the (observed) $0.1 - 2$ keV ROSAT band determined from the HRI count rate is about $2.9 \pm 0.25 \times 10^{45}\,{\rm erg\,s^{-1}}$, consistent with that inferred from the ASCA spectrum over the observed 0.5 − 2 keV band and a reasonable spectral extrapolation.

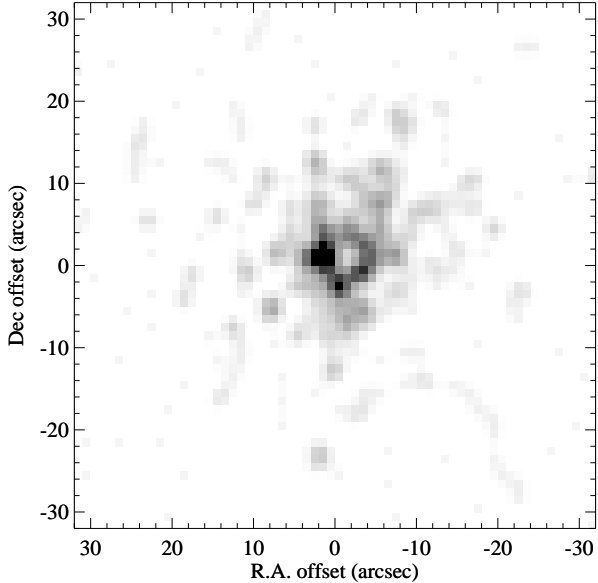

**Figure 1.** ROSAT HRI image of IRAS P09104+4109 binned into 1 arcsec pixels and smoothed with a gaussian of dispersion 1 arcsec. The centre of the hole is at 09h 10m 33.06s, +41d 08m 51.9d (J2000); the centre of the radio source (Hines & Wills 1993) lies to the NW of the hole on the surrounding bright ring. This small offset in consistent with the absolute pointing uncertainty of ROSAT ($\sim 10$ arcsec; David et al 1993).

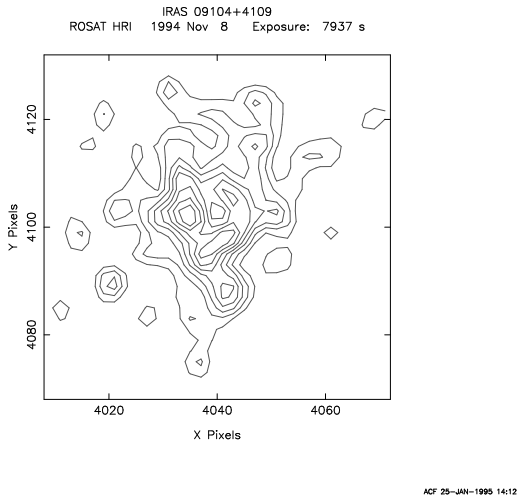

**Figure 2.** Contour map of the central part of the ROSAT HRI image of IRAS P09104+4109 shown in Fig. 1. The units on the axes are raw pixels each of 0.5 arsec. The contour levels are at 0.237, 0.355, 0.473. 0.592, 0.710, 0.828, 0.946 and 1.065 ct pixel$^{-1}$.

## 3 DISCUSSION

At least 90 per cent and, if the central hole in the image is real, probably all of the X-ray emission detected by ROSAT from IRAS P09104+4109 is from an extended source. The very similar X-ray luminosity to that detected with ASCA indicates that most of that emission too was from an extended source. Assuming that thermal bremsstrahlung dominates the emissivity, the gas density at $\sim 15$ arcsec radius (100 kpc) is $\sim 0.17\,\mathrm{cm}^{-3}$. The cooling time of the gas, $t_{\mathrm{cool}}$, assuming from the ASCA data a temperature of 11 keV (see below), is $\sim 4 \times 10^9$ yr and decreases inward. The peaked X-ray surface brightness profile and short cooling time are characteristic of a massive cooling flow in the host cluster about the central galaxy (for a review of cooling flows see Fabian 1994). The above figures indicate a mass cooling rate of $500\,\mathrm{M}_\odot\,\mathrm{yr}^{-1}$ within the inner 100 kpc and $\sim 1000\,\mathrm{M}_\odot\,\mathrm{yr}^{-1}$ within the $\sim 200$ kpc radius region of the cooling flow where $t_{\mathrm{cool}} < 10^{10}$ yr. The surface brightness profile, $S(r)$, of a typical cooling flow cluster varies as $S(r) \propto r^{-1}$ within $r_{\mathrm{cool}}$, and roughly as $S(r) \propto r^{-2}$ or steeper at larger radii. We compare a broken power-law with the observed profile in Fig. 3. Such steeply-peaked X-ray emission is not resolved by ASCA. CL 0016+16 has a much flatter X-ray profile with a core radius of 250 kpc (White, Silk & Henry 1981; see Fig. 3 for a comparison of a King model to the IRAS P09104+4109 data) and has no cooling flow. This explains why it is resolved by ASCA (Furazawa et al 1994).

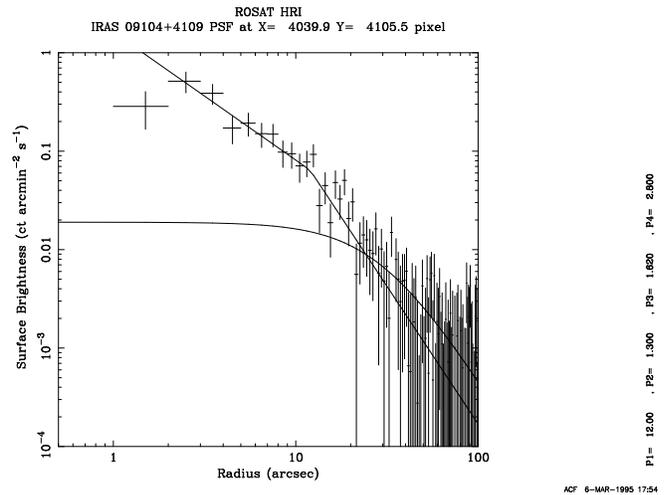

**Figure 3.** Surface brightness profile of IRAS P09104+4109 after background subtraction. The value inside 1 arcsec is effectively zero. Also shown are the best-fitting models over the range 2 − 60 arcsec of a broken power-law (the break is at $12^{+6}_{-4}$ arcsec, the slope $1.3 \pm 0.3$ at smaller radii and $2.8 \pm 0.4$ at larger radii) and of a King profile of core radius 30 arcsec and index 1.5.

We have refit the ASCA spectra with a model consisting of a cooling flow (see Johnstone et al 1992) and an isothermal gas (to represent the outer gas). Excess absorption (with 100 per cent covering fraction) is included on the cooling flow component (as found for many nearby cooling flows; White et al 1990; Allen et al 1993; Fabian et al 1994b) and Galactic absorption on both components. We obtain a reasonable fit ($\chi^2 = 571.7$ for 497 bins) for a gas temperature of $11.4^{+\infty}_{-3.2}$ keV, abundance $0.52 \pm 0.14 Z_\odot$, mass cooling rate $1003^{+202}_{-272}\,\mathrm{M}_\odot\,\mathrm{yr}^{-1}$ and excess column density $2.5^{+1.8}_{-1.1} \times 10^{21}\,\mathrm{cm}^{-2}$ (uncertainties are at the 90 per cent confidence level). The quality of the fit is slightly better than that for the absorbed power-law plus line model used previ-ously ($\chi^2 = 581$; Fabian et al 1994a). The (rest-frame) 0.5 – 10 keV luminosity of the source is $2.25 \times 10^{45}$ erg s$^{-1}$ with 70 per cent being in the cooling flow component. This high cooling flow fraction is a further reason that IRAS P09104+4109 was not resolved by ASCA.

We conclude that the ROSAT spatially-resolved image and the ASCA spectrum of IRAS P09104+4109 show that it is surrounded by a massive cooling flow. Indeed the central hole in the ROSAT image could be due to increased excess absorption within the innermost 10 kpc of the flow extinguishing the flux in the soft ROSAT band. (The mass of cold absorbing gas must then be $\sim 5 \times 10^{11}$ M$_\odot$.) It is also possible that the narrow optical emission-line nebulosity of the host galaxy is partly due to the cooling flow, since some of its properties such as extent and luminosity (although the gas appears more highly ionized) are similar to those seen in more nearby massive cooling flows (Crawford & Fabian 1992 and references therein).

Alternative interpretations of the central hole are that a) the radio-emitting plasma from the relatively bright central radio component has either heated or excluded the intracluster gas from that region, as observed in NGC 1275 (Böhringer et al 1993), or b) an outflow or eruption from the central AGN has excavated that region.

The X-ray appearance of IRAS P09104+4109 bears some striking similarities to those of 3C295, which is at a similar redshift ($z = 0.46$). Henry & Henriksen (1986) have presented the results of a 103 ks Einstein Observatory HRI observation of 3C295 and concluded that it too lies in a cooling flow. They estimated that the mass cooling rate is only $\dot{M} \sim 145$ M$_\odot$ yr$^{-1}$ by using the peaked luminosity of the inner region and a temperature of about $10^7$ K. The gas there is of course cooling from a much higher temperature of $\sim 10$ keV so $\dot{M} \sim 1000$ M$_\odot$ yr$^{-1}$ is a more reasonable estimate for 3C295. We note also that Henry & Henriksen neglected to convolve their King model with the PSF of the instrument, so concluded a minimum core radius for the extended component similar in size to the PSF.

Other, similarly large, cooling flows at moderate redshifts are Zw3146 ($\dot{M} \sim 1000$ M$_\odot$ yr$^{-1}$, $z = 0.3$; Edge et al 1994), and A1068 ($\dot{M} \sim 400$ M$_\odot$ yr$^{-1}$, $z = 0.14$; Allen et al 1995) which is also infrared luminous (although 30 times less so than IRAS P09104+4109; Allen 1995). Both of these were found from the ROSAT All-Sky Survey (RASS) data (Allen et al 1992). Distant, cooling-flow-dominated clusters such as IRAS P09104+4109 may not appear extended in the RASS.

The iron abundance of IRAS P09104+4109 of $\sim 0.5$ is unusually high for such a high temperature cluster (see e.g. correlations in Fabian et al 1994c). An abundance of about half that value would be more typical for nearby clusters of that luminosity. Perhaps the iron abundance was higher at earlier times, due say to an abundance gradient such as is now seen in the low temperature Centaurus cluster (Fukazawa et al 1994). We note however that the high temperature distant cluster A370, which has no large cooling flow, also has a high abundance of 0.5 (Bautz et al 1994). It remains possible that part of the line in P09104+4109 is due to resonance scattering and fluorescence of the innermost cooling flow within a kpc of the nucleus, as proposed in our earlier work (Fabian et al 1994a). The hard continuum can certainly not be electron scattered by gas at larger radii without the thermal emission from the hot electrons outshining the scattered flux (for a reasonable AGN luminosity comparable with the IR value). The optical depth for resonance scattering is however much larger than the Thomson optical depth and some contribution is likely. Future observations of the iron line at high spatial resolution with AXAF are required to separate such components.

IRAS P09104+4109 is the first hyperluminous IRAS galaxy to be found in a cooling flow. Part of the infrared and much of the optical line luminosities may be due to the flow but the detection of broad MgII emission and strong optical polarization by Hines & Wills (1993) indicate a central AGN. The gas deposited by the flow is an obvious source of fuel for such an AGN and they may be linked by Compton cooling of the innermost flow, as discussed by Fabian & Crawford (1990). Further work is necessary to unravel the different components of this interesting object.

## 4 ACKNOWLEDGEMENTS

We thank Chris Reynolds for help. The reduction was carried out using XIMAGE and we thank L. Angelini and N. White for assistance. CSC and ACF thank the PPARC and Royal Society for help, respectively.